\documentclass[twocolumn,showpacs,amssymb,aps,nofootinbib,floatfix]{revtex4-1}
\usepackage[latin1]{inputenc}
\usepackage{bm}
\usepackage{epsfig}
\usepackage{amsthm}
\usepackage{amsfonts}
\usepackage{float}
\usepackage{amsmath,amssymb}
\usepackage{color}
\usepackage{graphicx}
\usepackage{dcolumn}
\usepackage{bm}
\def\nn{\nonumber}
\newcommand{\beq}{\begin{equation}}
\newcommand{\eeq}{\end{equation}}
\newcommand{\beqa}{\begin{eqnarray}}
\newcommand{\eeqa}{\end{eqnarray}}
\newcommand{\lef}{\left(}
\newcommand{\rig}{\right)}
\newcommand{\la}{\langle}
\newcommand{\ra}{\rangle}

\begin{document}

\title{Pseudorapidity profile of transverse momentum fluctuations in heavy ion collisions}
\author{Sandeep Chatterjee}
\email{Sandeep.Chatterjee@fis.agh.edu.pl}
\author{Piotr Bo{\.z}ek}
\email{piotr.bozek@fis.edu.agh.pl}
\affiliation{AGH University of Science and Technology,\\ 
Faculty of Physics and Applied Computer Science,\\
al. Mickiewicza 30, 30-059 Krakow, Poland}

\begin{abstract}
We investigate pseudorapidity correlations of the average transverse flow of particles emitted in relativistic heavy-ion collisions.
We employ 3+1 dimensional viscous relativistic hydrodynamics with initial conditions from 
the quark Glauber Monte Carlo  model to confront the recent measurements  
on the pseudorapidity correlations of the transverse momentum fluctuations in Pb+Pb 
collisions at $\sqrt{s_{NN}}=2760$~GeV. We find good agreement between the model predictions 
and data. Further, we study two other observables build with the covariance 
of the average transverse momentum in different rapidity bins. These
 observables have better 
stability under
various systematics, thus allowing for a robust comparison between data
 and model.  The transverse flow-transverse flow correlation coefficient
is directly related to correlations of the underlying collective flow at different pseudorapidities. The 3-bin measure of $p_T$ factorization breaking in pseudorapidity gives an estimate of possible decorrelation of the average transverse flow 
in the
longitudinal direction. 
\end{abstract}


\maketitle

\section{Introduction}\label{sec.intro}

Relativistic heavy ion collisions produce a hot and dense strongly interacting fireball that 
rapidly expands and freezes into free-streaming hadrons that are registered by the detectors. 
There has already been enormous progress in understanding the transverse (to the 
beam direction)  evolution of the initial transverse spatial anisotropy of the overlap 
zone into the final state momentum anisotropy through equations of viscous relativistic 
hydrodynamics \cite{Heinz:2013th,*Gale:2013da,*Ollitrault:2010tn}. An 
analogous effort to comprehend the longitudinal dynamics has picked up pace as well. It is widely 
believed on the basis of causality arguments that correlations over large rapidity gaps arise 
due to early time dynamics. Thus a good knowledge of the rapidity profile of the fireball 
will lead to a better understanding of the initial conditions and
 early dynamics in 
the longitudinal direction.

There have been several studies on pseudorapidity correlations of event-plane orientations  both in theory~\cite{Bozek:2010vz,Petersen:2011fp,Xiao:2012uw,Jia:2014vja,Pang:2014pxa}
 as well as in experiments~\cite{Khachatryan:2015oea,ATLAS-CONF-2017-003}. 
 Further, multiplicity correlation at different rapidities have been studied as well~\cite{Back:2006id,*Abelev:2009ag,*ATLAS:2015kla}. 
 However, the latter  is more challenging 
as it is difficult to disentangle different sources of short and long range 
correlations.

Fluctuations and correlation of transverse momentum are believed to give insight into the mechanism of energy deposition 
in heavy ion collisions and could serve as a probe of the properties of the medium formed
\cite{Gazdzicki:1992ri,Stodolsky:1995ds,Shuryak:1997yj,Mrowczynski:1997kz,Liu:1998xf,%
Voloshin:1999yf,Baym:1999up,Voloshin:2001ei,Korus:2001au,DiasdeDeus:2003ei,%
Broniowski:2005ae,Sharma:2008qr,Hama:2009pk,Trainor:2015swa,Basu:2016ibk,Gavin:2016nir,Liu:2016apq,Bzdak:2017shg}.
Collective expansion of the fireball can be at the origin of fluctuations
 of the average transverse momentum of particles emitted 
in an event \cite{Broniowski:2009fm,Bozek:2012fw,Mazeliauskas:2015efa}.
In events where the initial transverse size of the fireball is larger, 
a smaller transverse flow is generated. Experimentally,
 fluctuations of the average transverse momentum have be analyzed in 
heavy-ion and proton-proton collisions
\cite{Adams:2003uw,Adamova:2003pz,Adler:2003xq,Anticic:2003fd,Abelev:2014ckr}.
Hydrodynamic models with Monte Carlo Glauber initial conditions predict 
 sizable transverse momentum fluctuations, in semiquantitative agreement 
with the data.

In this paper we study correlations of the average
 transverse momentum in different pseudorapidity bins.
The decorrelation of the average transverse momentum in two pseudorapidity bins 
is a measure of the  factorization breaking of the flow. In 
the hydrodynamic model such a factorization breaking is possible due to event-by-event fluctuations.
The ALICE collaboration 
has reported preliminary measurement of pseudorapidity correlations
 of   average transverse momenta at different pseudorapidities 
 \cite{ptposterQM}.
The  correlation of the average 
transverse momenta has been also  calculated  in some  event generators
\cite{Kovalenko:2016bcx}. 
We calculate this observable in the hydrodynamic framework and find a fair 
agreement with data. 

In order to reduce strong statistical fluctuations we propose to measure a
modified correlation coefficient between transverse momenta 
in different pseudorapidity bins, removing the self-correlation contribution
 to the variance
\cite{Adams:2003uw}. This modified definition could be directly 
interpreted as a measure of  transverse flow-transverse flow 
In order to reduce  non-flow contributions a 3-bin measure of the transverse
 momentum decorrelation can be used, in analogy to the one used for
 the azimuthal flow coeffcients
\cite{Khachatryan:2015oea}. In the hydrodynamic model with Monte Carlo Glauber
initial conditions, the decorrelation of the collective average 
transverse momentum is 0.8-1.0 for pseudorapidities $|\eta|<2.5$. Additional
 non-flow effects in experimental estimates  can modify  these values.

\section{Glauber model and viscous hydrodynamics}\label{sec.model}

We use event-by-event viscous relativistic  hydrodynamics 
with fluctuating initial conditions \cite{Schenke:2010rr}.
The initial conditions are take from a Monte Carlo Glauber model. We use
 the quark Glauber model, where each nucleon is composed of three quarks
( details and parameters can be found in \cite{Bozek:2016kpf}).
The Monte Carlo Glauber model predicts for each event the positions 
of the participant quarks. The initial entropy density is deposited at the participant positions 
in the form of Gaussians $g_i(x,y)$ of width $0.3$~fm in the transverse plane.

The longitudinal profile used in our study involves asymmetric deposition
in space-time rapidity $\eta_{\parallel}$ from  left  or right 
 going participants. Event-by-event fluctuations in the number of left and 
right going participants ($N_-$ and $N_+$ respectively) or their positions leads to a 
decorrelation of the flow and entropy at  different rapidities
\begin{equation}
s(x,y,\eta_\parallel)=\sum_{i=1}^{N_+} g_i(x,y) f_+(\eta_+) + \sum_{i=1}^{N_-} g_i(x,y) f_-(\eta_\parallel)   \ .
\end{equation} 
 The longitudinal density profile is 
\begin{equation}
f_{\pm}(\eta_\parallel)= \frac{y_{beam}\pm \eta_\parallel}{2 y_{beam}} H(\eta_\parallel)\  \mbox {for } \ |\eta_\parallel|<y_{beam} \ ,
\label{eq:lprof}
\end{equation}
and \cite{Hirano:2002ds}
\begin{equation}
H(\eta_\parallel)=exp\left(-\frac{(|\eta_\parallel|-\eta_p)^2\Theta(|\eta_\parallel|-\eta_p)}{2\sigma_\eta^2}\right) ,
\end{equation}
where $\sigma_\eta=1.4$ and $\eta_p=1.15$.
Model calculations  presented below illustrate the application of the transverse momentum correlation and do not represent an extensive study of
 possible scenarios for fluctuations in the initial state or in the dynamics. 
We note that 
in other models of the initial state the initial decorrelation 
in space-time rapidity may be larger  \cite{Petersen:2011fp,Xiao:2012uw,Bozek:2015bna,Monnai:2015sca}.
The initial fireball expands collectively. 
We set the shear viscosity to entropy ratio to $0.08$. At the freeze-out 
temperature of $150$~MeV  statistical emission of particles occurs
 \cite{Chojnacki:2011hb} (particlization procedure). 
This framework predicts correctly the fluctuations of the average 
transverse momentum in Pb+Pb collisions \cite{Bozek:2017elk}, which is a good starting point to investigate $p_T$ correlations in pseudorapidity. 

Generated 
events have realistic
 multiplicities and involve non-flow correlations from resonance decays. 
Alternatively, we generate events with oversampled multiplicity for each
freeze-out hyperspace. In such events non-flow effects are reduced and
 one measures directly fluctuations and correlations of the average transverse
 flow in the spectra.

\section{Pearson correlation coefficient}\label{sec.observable}

 In this study, we focus on the statistical properties 
of the average transverse momentum in the event $[p_T]$
\beqa
 [p_T] &=& \frac{1}{n}\sum_i p_i. \label{eq.pT}
\eeqa
where $n$ is the number of particles in the  event in a given
 acceptance window ($p_i$ is the transverse momentum of particle $i$). 
Thus, $[...]$ denotes 
averaging over the particles in an event. Similarly, the event average $\la \dots  \ra $ of $[p_T]$ 
could be defined as
\beq
 \la[p_T]\ra = \frac{1}{N_{ev}}\sum_{ev} [p_T] \label{eq.pTensembleaverage}
\eeq
where $N_{ev}$ is the total number of events in the ensemble. 
We  look at the correlation of $[p_T]$ in different pseudorapidity bins. The 
average transverse momentum is calculated for $p_T$ in the range $0.2-2.0$~GeV.
The observable 
$b\lef [p_T]_F, [p_T]_B \rig$ that was measured by the ALICE collaboration is defined as~\cite{ptposterQM}
\begin{widetext}
\beqa
 b\lef [p_T]_F,[p_T]_B\rig  &=& \frac{\text{Cov}\lef [p_T]_F,[p_T]_B\rig}
 {\sqrt{\text{Var}\lef[p_T]_F\rig}\sqrt{\text{Var}\lef[p_T]_B\rig}}\nn\\
 &=& \frac{\la\lef[p_T]_{F}-\la[p_T]\ra_F\rig \lef[p_T]_{B}-\la[p_T]\ra_B\rig\ra}
 {\sqrt{\la\lef[p_T]_{F}-\la[p_T]\ra_F\rig^2\ra}\sqrt{\la\lef[p_T]_{B}-\la[p_T]\ra_B\rig^2\ra}} \nn \\
& = & \frac{\la \frac{1}{n_F}\sum_{i\in F} (p_i- \la [p_T] \ra_F) \frac{1}{n_B}\sum_{j\in B} (p_j -\la [p_T] \ra_B)\ra }
{\sqrt{\la \frac{1}{n_F^2} \sum_{i,j \in F} ( p_i-\la[p_T]\ra_F )( p_j-\la[p_T]\ra_F )\ra}\sqrt{\la \frac{1}{n_B^2} \sum_{i,j \in B} ( p_i-\la[p_T]\ra_B )( p_j-\la[p_T]\ra_B )\ra}}\label{eq.bcorr}
\eeqa
where $F\in[\eta_{min},\eta_{max}]$ and $B\in[-\eta_{max},-\eta_{min}]$ denote the two 
pseudorapidity bins in the forward(F) and backward(B) hemispheres whose correlation strength is being 
studied. Mathematically, the quantity  $b\lef [p_T]_F,[p_T]_B\rig$
 represents the Pearson correlation coefficient between $[p_T]_F$ and $[p_T]_B$
averaged over the ensemble of events.
\end{widetext}

 \begin{figure}
\vspace{-4mm}
 \hspace{-9mm} 
  \includegraphics[scale=0.45]{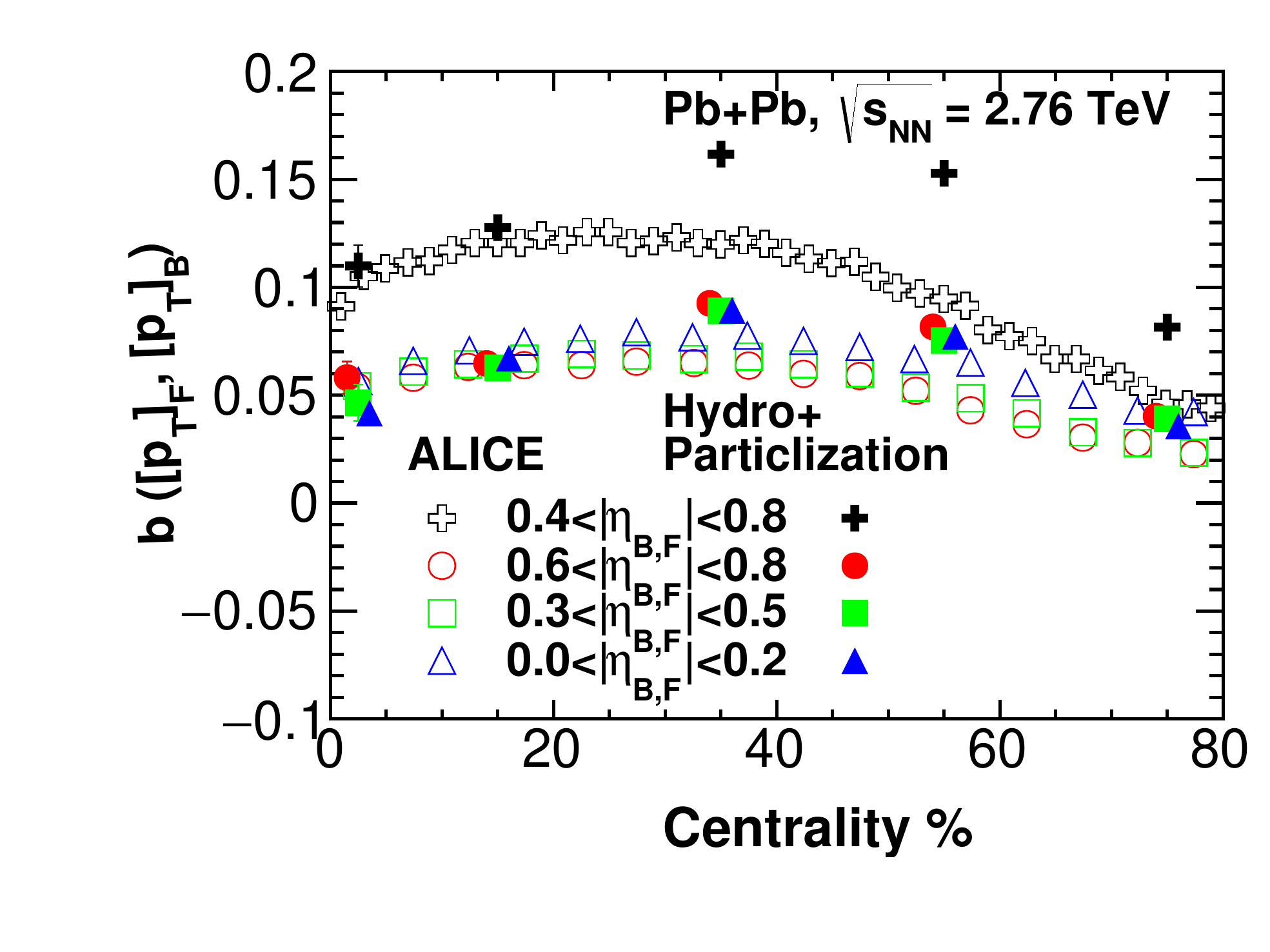}
\vspace{-6mm}
  \caption{(Color online) Preliminary data from ALICE Collaboration~\cite{ptposterQM} on Pearson correlation coefficient  $b\lef [p_T]_F,[p_T]_B\rig$ in Pb+Pb 
collisions at $\sqrt{s}=2760$~GeV  are compared to hydrodynamic model predictions. Results are shown for different choices of pseudorapidity bins where the average transverse momentum is calculated. } 
  \label{fig.alice}
 \end{figure}

\begin{figure}
\vspace{-4mm}
 \hspace{-9mm} 
\includegraphics[scale=0.45]{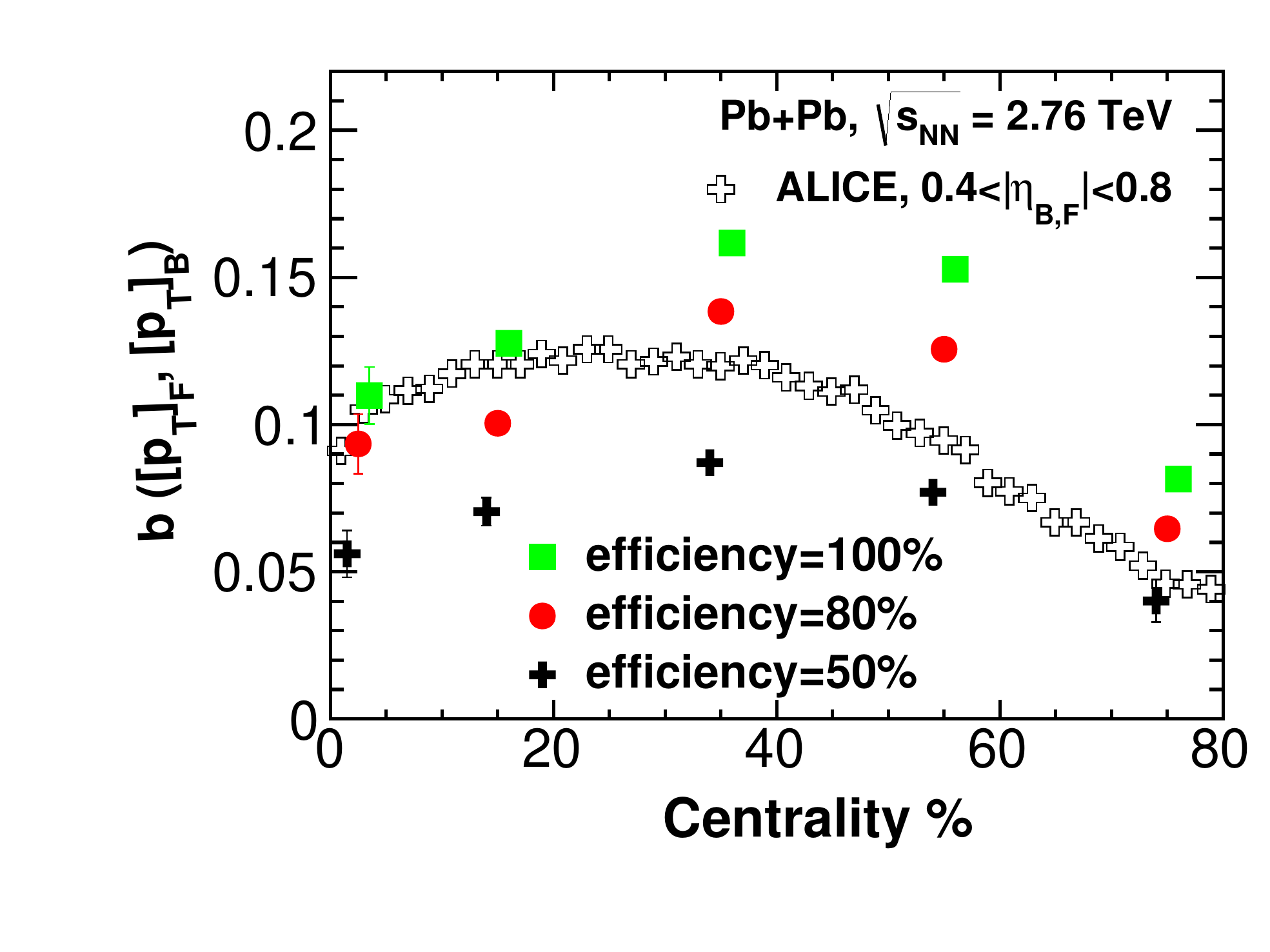}
\vspace{-6mm}
  \caption{(Color online) Preliminary data from ALICE Collaboration~\cite{ptposterQM} on the correlation coefficient   $b\lef [p_T]_F,[p_T]_B\rig$ in
 two pseudorapidity 
bins ($0.4<|\eta_{F,B}|<0.8$) in Pb+Pb collisions at $\sqrt{s}=2760$~GeV. Hydrodynamic model  results are shown for   $100\%$  (squares) , $80\%$ (circles) and $50\%$ efficiency (crosses).  } 
  \label{fig.eff}
 \end{figure}

In Fig. \ref{fig.alice} are shown preliminary results for the Pearson correlation coefficient  $b\lef [p_T]_F,[p_T]_B\rig$ measured in Pb+Pb collisions at different centralities.
The  centrality dependence of the correlation coefficient is
 non-monotonous, the correlation is maximal for semi-central collisions. 
Interestingly the calculation in the hydrodynamic model with quark Glauber initial conditions (filled  symbols in Fig. \ref{fig.alice}) reproduces fairly well the magnitude of the correlation and quantitatively it shows a similar centrality dependence. The correlation coefficient  $b\lef [p_T]_F,[p_T]_B\rig$ shows a  strong dependence
on the width of the pseudorapidity bin used. On the other hand, the dependence on the separation of the two bins is weak. 
It is difficult to interpret the experimental data for this quantity as the result depends strongly on the chosen  bin width.

As shown in Fig. \ref{fig.eff} the results depend also on the efficiency. 
A change from of the efficiency from $100\%$ to $80\%$ or $50\%$ in the model calculation 
shifts the value of  $b\lef [p_T]_F,[p_T]_B\rig$ significantly. The reason for this behavior is
a strong dependence of the variance $Var([p_T])$ in Eq. (\ref{eq.bcorr}) 
on multiplicity. As argued in Ref. \cite{Adams:2003uw} this quantity contains
a contribution from statistical fluctuations in the sampling of transverse 
 momenta from a spectrum. It cannot be a measure of the fluctuations of the 
transverse flow. This effect is especially important when using small bin 
widths.

\section{Transverse flow-transverse flow  correlation coefficient}
\label{sec:flowflow}

The variance of the average transverse momentum of a spectrum
can be estimated when using a formula excluding self-correlations
\beq
C_{p_T}=\frac{1}{n(n-1)}\sum_{i\neq j} (p^i_T -[p_T])(p^j_T-[p_T]) 
\label{eq:cpt}
\eeq
For independent particle emission, this formula eliminates statistical 
fluctuations from the estimate of the  variance of the transverse flow. 
It has been used
 in experimental measurements of the $p_T$ fluctuations \cite{Adams:2003uw,Adamova:2003pz,Adler:2003xq,Anticic:2003fd,Abelev:2014ckr} 
and to define the transverse momentum-flow 
correlation coefficient \cite{Bozek:2016yoj}.

 \begin{figure}
\vspace{-4mm}
 \hspace{-9mm} 
 \includegraphics[scale=0.45]{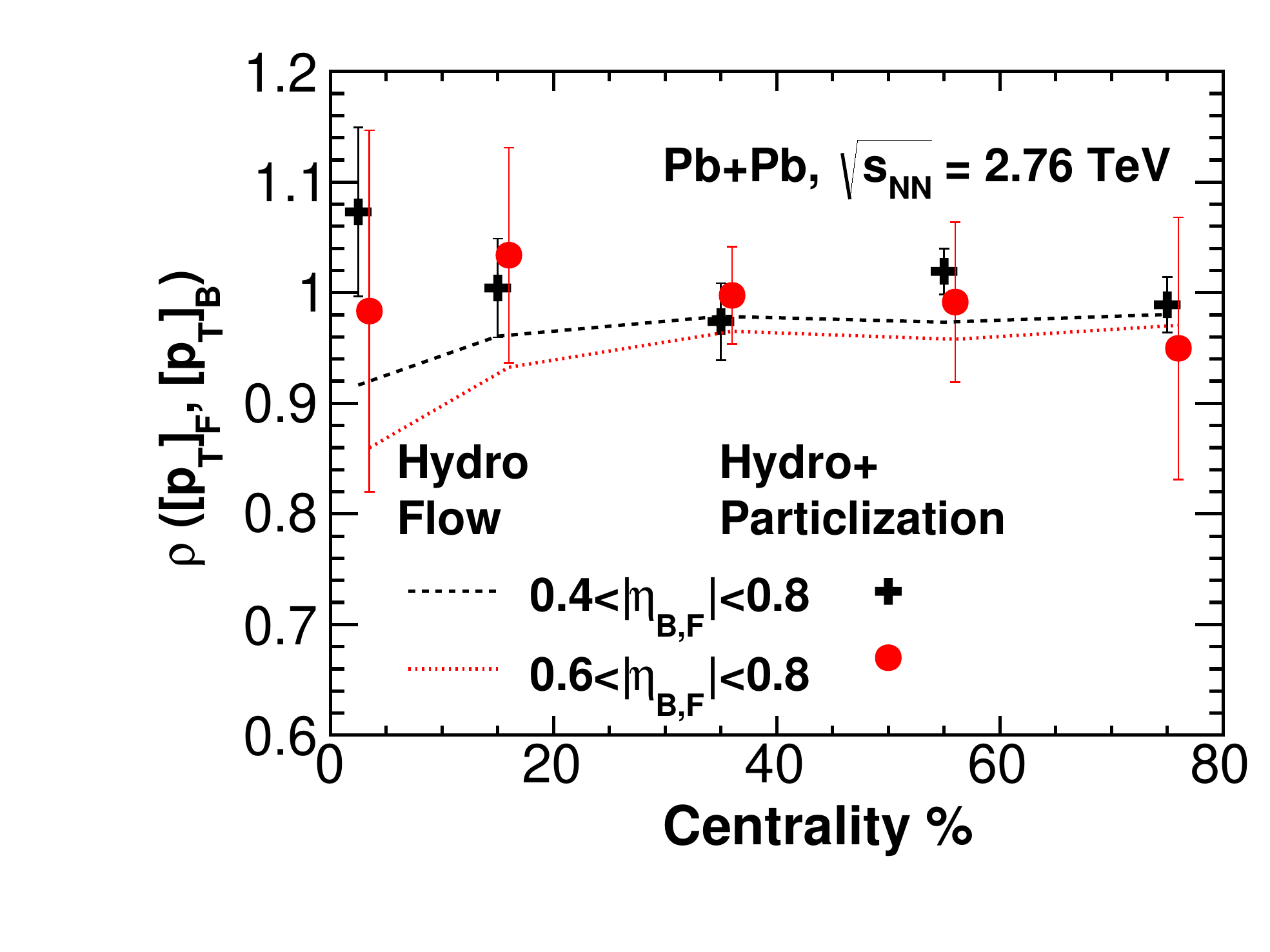} 
\vspace{-6mm}
  \caption{(Color online) Transverse flow-transverse flow correlation coefficient $\rho\lef [p_T]_F,[p_T]_B\rig$ predicted in the hydrodynamic model plotted as a function of collision centrality. Results are shown for two choices of pseudorapidity intervals $0.4<|\eta_{B,F}|<0.8$ (crosses and dashed line) and $0.6<|\eta_{B,F}|<0.8$ (circles and dotted line). Symbols represent the results obtained from events generated with realistic multiplicities and non-flow correlations from
 resonance decays, while the lines are obtained by integrating the particle spectra and include only correlations from flow.} 
  \label{fig.rhocent}
 \end{figure}

The correlation coefficient for the average transverse flow in two 
pseudorapidity bins takes the form
\beq
 \rho\lef [p_T]_F,[p_T]_B\rig = \frac{\text{Cov}\lef [p_T]_F,[p_T]_B\rig}
 {\sqrt{C_{p_T}^F C_{p_T}^B}} \label{eq:rho}
\eeq
where self-correlations are excluded in the calculation of the variances
 in the denominator. If only flow correlations are present, the correlation coefficient (\ref{eq:rho}) is smaller than $1$. As discussed for  the factorization breaking of azimuthal flow coefficients, a value larger than $1$ signals a significant contribution from non-flow correlations \cite{Gardim:2012im}.

Fig. \ref{fig.rhocent} presents the results for the transverse flow-transverse flow correlation coefficient as a function of centrality 
for the same kinematic range 
as the observable  $b\lef [p_T]_F,[p_T]_B\rig$ discussed in the previous section. The striking result is that $\rho(\lef [p_T]_F,[p_T]_B\rig$ is very different
 from  $b\lef [p_T]_F,[p_T]_B\rig$. The correlation coefficient $\rho$ is
close to $1$,  $8-10$ times larger than  $b\lef [p_T]_F,[p_T]_B\rig$. The only 
difference 
between the two observables is in the definition of the variance of the transverse momentum. 
In the limit of infinite multiplicity both quantities should be equal. 
In the hydrodynamic model,  genuine correlations of the collective 
transverse flow can be
calculated using  oversampled events.  The results are shown with lines in Fig. \ref{fig.rhocent}.  The correlation coefficient for 
the collective transverse  flow is always smaller than $1$. The factorization
 breaking of the average transverse flow in our hydrodynamic calculation
 is small;
$\rho(\lef [p_T]_F,[p_T]_B\rig > 0.8$.
We notice that the decorrelation is stronger when  the separation between the two pseudorapidity bins increases. In realistic events non-flow correlations from 
resonance decays are present. These non-flow correlations increase the value of $\rho(\lef [p_T]_F,[p_T]_B\rig$ with respect to the values corresponding 
to collective flow only. However, no definite conclusions on deviations of the correlation coefficient from $1$ can be made with
 the statistics available in our simulations.

 \begin{figure}
\vspace{-4mm}
\hspace{-9mm}
  \includegraphics[scale=0.45]{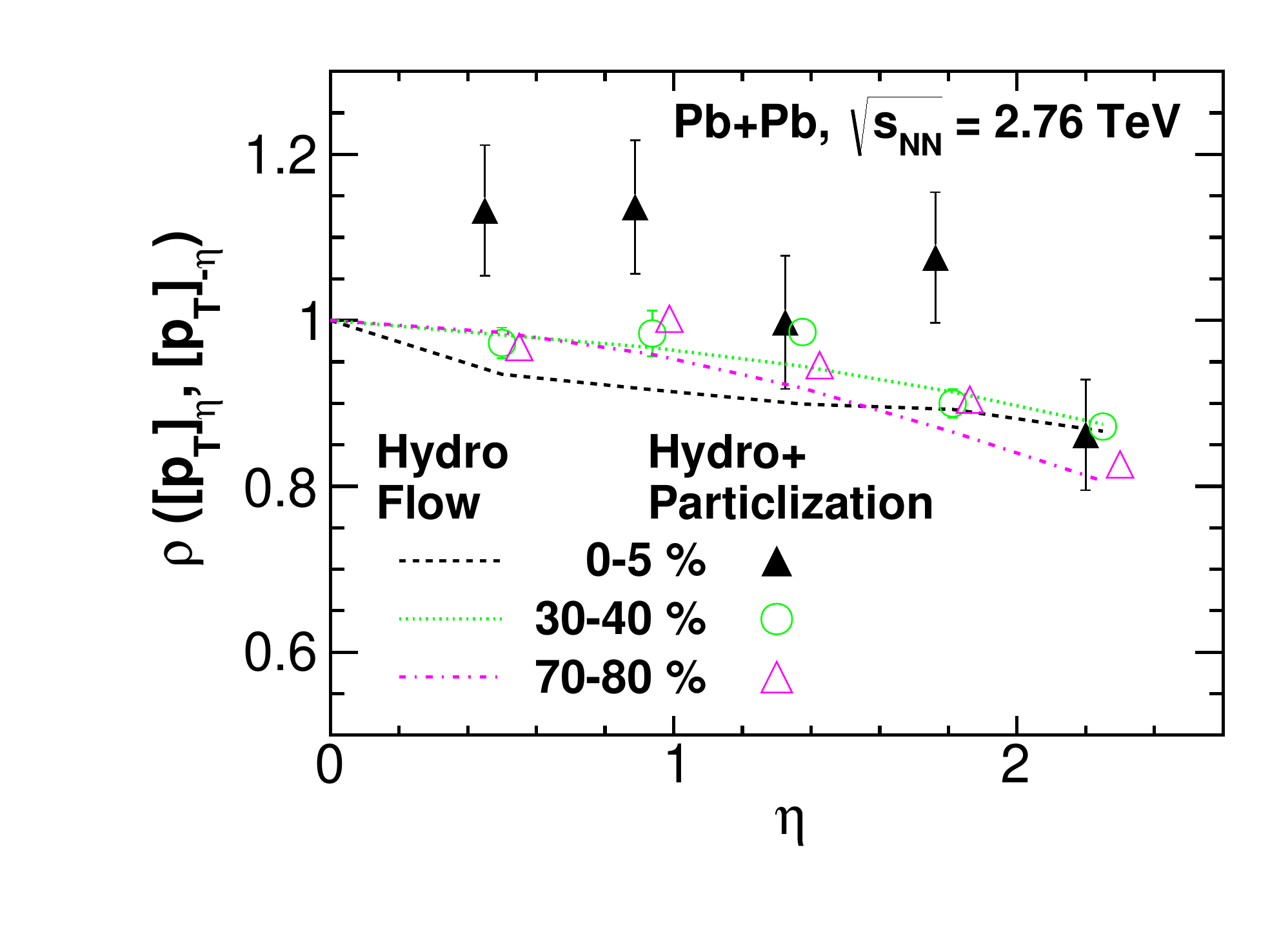}
\vspace{-6mm}
  \caption{(Color online) Transverse flow-transverse flow correlation coefficient $\rho\lef [p_T]_\eta,[p_T]_{-\eta}\rig$ plotted plotted as a function of $\eta$ for three different centralities. Symbols are for results from realistic finite 
multiplicity events and lines are obtained from spectra integration (see caption of Fig. \ref{fig.rhocent}). } 
  \label{fig.rho}
 \end{figure}

The correlation coefficient can be calculated for larger pseudorapidity intervals. We move the two pseudorapidity bins in the range $|\eta|<2.5$.
The results for the  correlation coefficient 
$\rho(\lef [p_T]_\eta,[p_T]_{-\eta}\rig$ as a function of the bin position $\eta$
 are shown in Fig. \ref{fig.rho} for three different centralities. For centralities $30-40\%$ and $70-80\%$ the decorrelation of the average transverse 
momenta deviates from $1$ for bin separation $2\eta>3$. It indicates that 
the decorrelation of the transverse flow could be observed in experiments 
using bins with a large pseudorapidity separation. With sufficient statistics,
 non-flow effects could be estimated by comparing results using 
same- and opposite-charged pairs.

  \begin{figure}
\vspace{-4mm}
\hspace{-9mm}
  \includegraphics[scale=0.45]{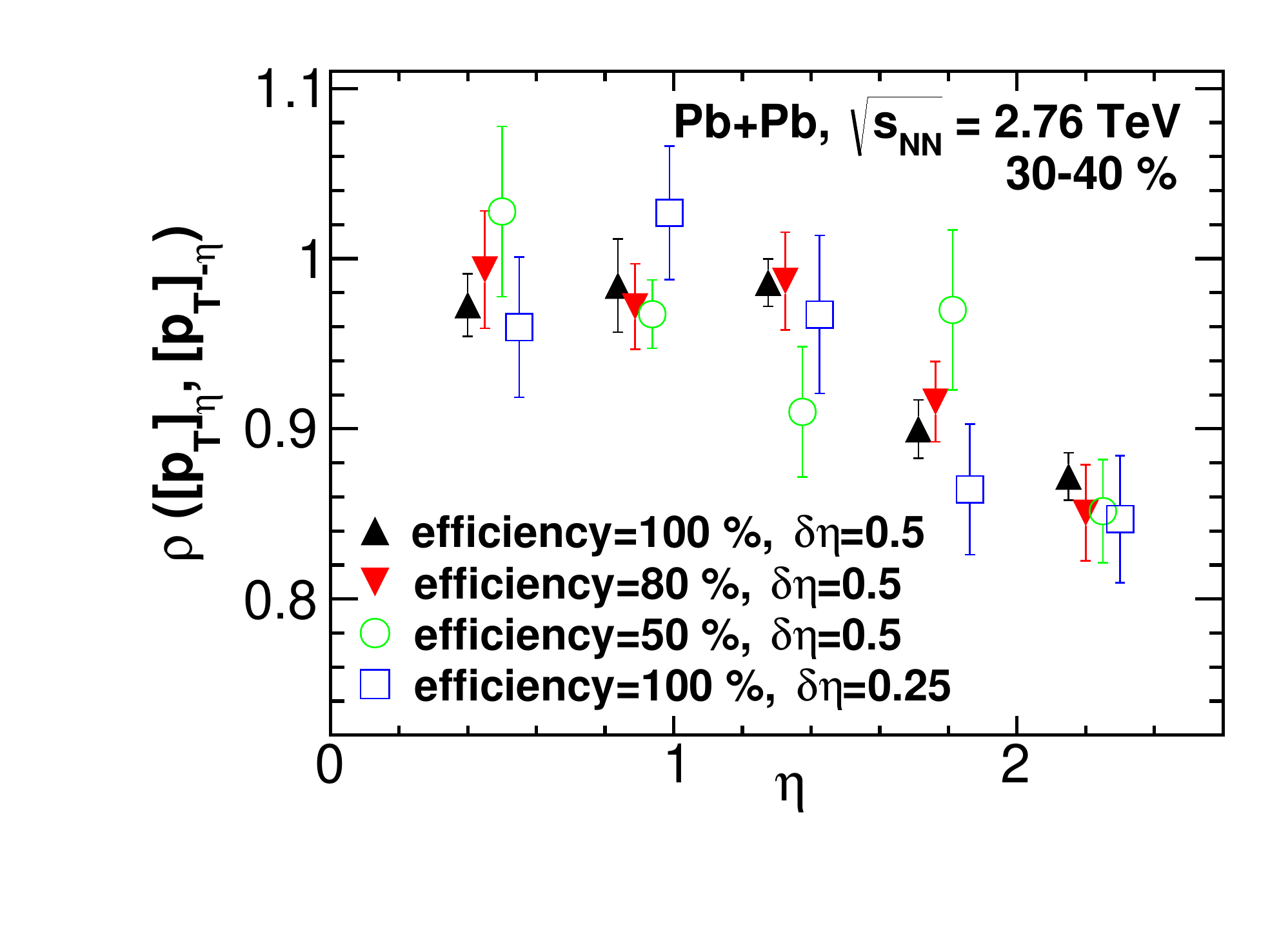}
\vspace{-6mm}
  \caption{(Color online) Transverse flow-transverse flow correlation coefficient $\rho\lef [p_T]_\eta,[p_T]_{-\eta}\rig$  plotted as a function of $\eta$. We compare results of hydrodynamic calculations with assumed efficiency $100\%$, $80\%$, and $50\%$, all with bin width $\delta \eta =0.5$. Squares denote the results obtained with bin of width $0.25$ and efficiency $100\%$.} 
  \label{fig.rhoeff}
 \end{figure}

We illustrate the robustness of the definition of the
 transverse flow-transverse flow correlation coefficient  (Eq. \ref{eq:rho}) 
by comparing results obtained for 
 different effciencies and bin widths. As shown in Fig. \ref{fig.rhoeff} the correlation 
coefficients obtained using $100\%$, $80\%$, and $50\%$ effciencies are
compatible within statistical errors. Further, the results from $\eta$ bin widths of $0.5$ 
and $0.25$ also agree with each other. Thus, the fluctuations of transverse flow calculated 
without self-correlations (Eq. \ref{eq:cpt}) have the desired property that the measured values 
do not depend on the multiplicity of the subsample used for the calculation.

  \begin{figure}
\vspace{-4mm}
\hspace{-9mm}
  \includegraphics[scale=0.45]{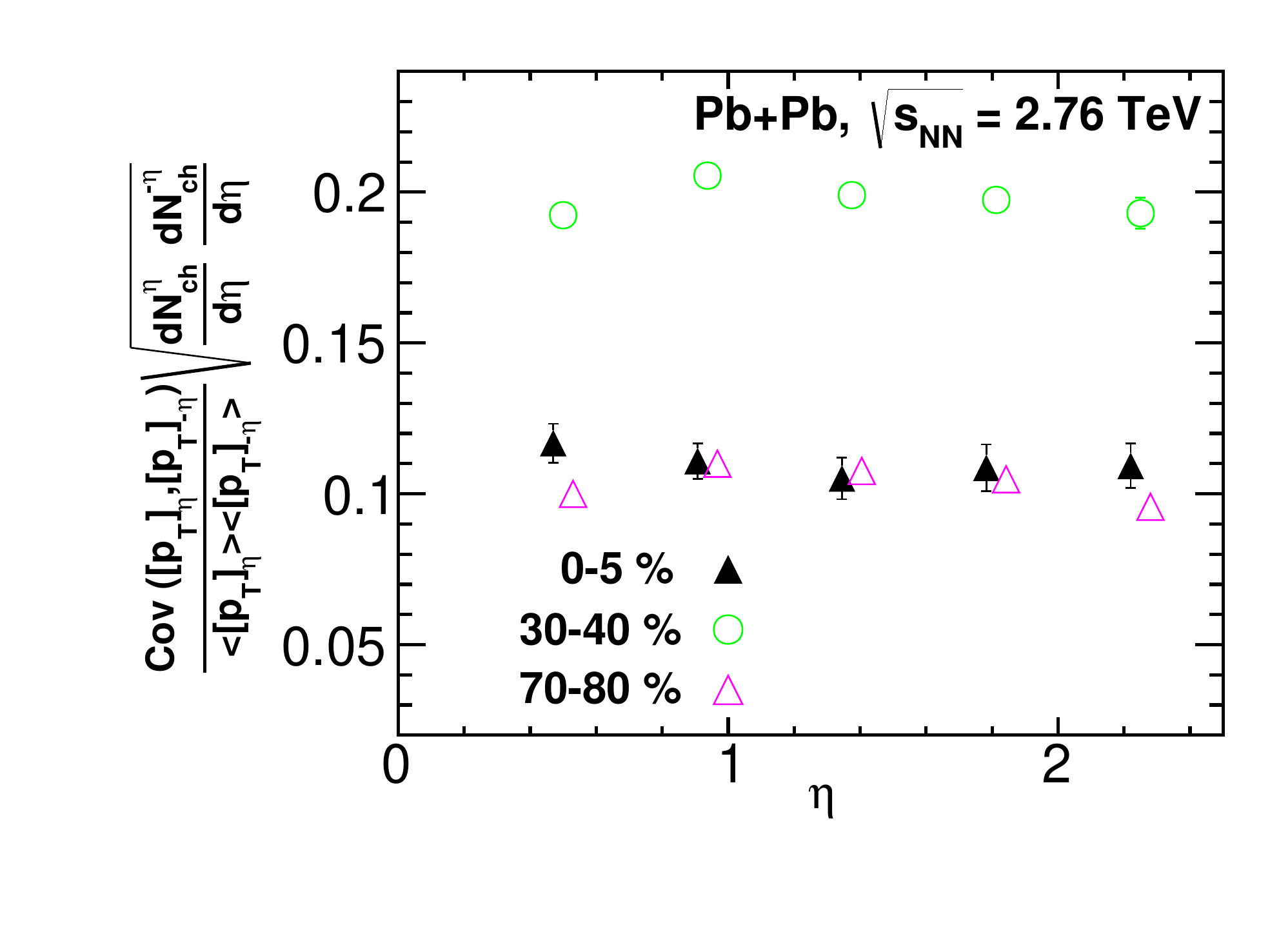}
\vspace{-6mm}
  \caption{(Color online) Covariance $\text{Cov}\lef [p_T]_\eta,[p_T]_{-\eta}\rig$ of the average transverse momentum in two 
pseudorapidity bins scaled by the product of  the square roots of the charged particle densities and by the inverse of the average transverse momenta  in the two pseudorapidity bins. Results are shown as a function of the bin position  $\eta$, for three different centralities.}
  \label{fig.covsca}
 \end{figure}

The average transverse momentum in  a pseudorapidity bin is an intensive 
quantity. It suggests to define the scaled covariance of the 
average transverse momenta as
\beq
\frac{\text{Cov}\lef [p_T]_\eta,[p_T]_{-\eta}\rig}{\la [p_T] \ra_\eta \la [p_T] \ra_{-\eta}} \ .
\eeq
The above quantity shows a very strong centrality dependence, similar to the scaled transverse momentum fluctuations \cite{Abelev:2014ckr,Bozek:2017elk}
\beq
\frac{\sqrt{C_{p_T}}}{\la [p_T] \ra} \simeq \left( \frac{dN}{d\eta} \right)^{-1/2} \ .
\label{eq:scaling}
\eeq
The above scaling is expected if transverse momentum fluctuations come from 
a superposition of independent sources. We note that experimental
 data and hydrodynamic
model results show systematic deviations from this trend.
In Fig. \ref{fig.covsca} is shown the scaled variance
\beq
\frac{\text{Cov}\lef [p_T]_\eta,[p_T]_{-\eta}\rig}{\la [p_T] \ra_\eta \la [p_T] \ra_{-\eta}} \sqrt{\frac{dN^\eta}{d\eta}\frac{dN^{-\eta}}{d\eta}}
\label{eq:covsc}
\eeq
for three different centralities. One notices a weak dependence of this quantity on the separation of   pseudorapidity bins. 
The remaining  dependence on centrality of the magnitude of the scaled covariance can be explained by the deviation of the
transverse momentum fluctuations from the scaling Eq. (\ref{eq:scaling}).
 Unlike the correlation coefficient $\rho\lef [p_T]_\eta,[p_T]_{-\eta}\rig$, the scaled covariance (\ref{eq:covsc}) can be 
 calculated without involving the variance of the transverse flow. However, there is no  simple physical interpretation of 
 this quantity in the hydrodynamic model. Moreover, it is difficult to disentangle the dependence on the bin separation in 
 pseudorapidity of the covariance from those of the average transverse momentum and charged particle density.

\section{3-bin decorrelation measure}
\label{sec:3bin}

The transverse flow-transverse flow 
 correlation coefficient  (\ref{eq:rho}) is plagued with
non-flow contributions. The non-flow contribution to the covariance 
is reduced for widely separated pseudorapidity bins. On the other hand, 
 the   $p_T$ fluctuations  measure $C_{p_T}$  has always a contribution 
from non-flow correlations.
The CMS Collaboration has proposed to use a 3-bin decorrelation measure for 
azimuthal flow coefficients \cite{Khachatryan:2015oea}. A similar
decorrelation for transverse flow correlations in pseudorapidity reads
\beq
r_{p_T}(\eta)=\frac{\text{Cov}\lef [p_T]_F,[p_T]_{-\eta}\rig}{\text{Cov}\lef [p_T]_F,[p_T]_{\eta}\rig} 
\ .
\label{eq:3bin}
\eeq
For Pb+Pb collisions we  symmetrize  the above 
 expression in 
forward-backward directions 
  to increase statistics. The forward bin is taken as $2.0<\eta<2.5$ and the center of the  second
bin of width $0.5$ is moved from $0.1$ to $1.6$.
  
  \begin{figure}
\vspace{-4mm}
\hspace{-9mm}
  \includegraphics[scale=0.45]{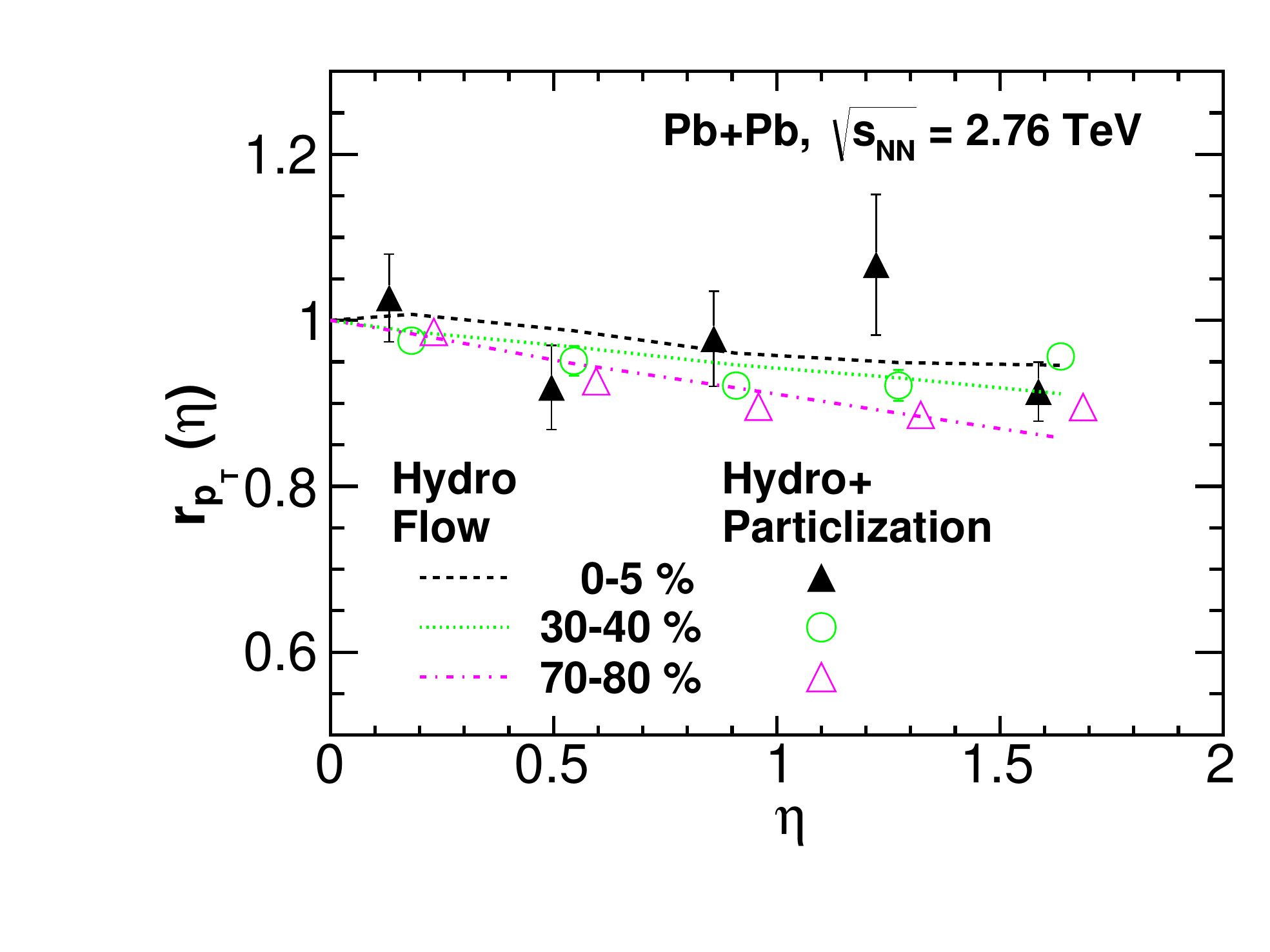}
\vspace{-6mm}
  \caption{(Color online) Factorization breaking coefficient $\text{r}_{p_T}\lef\eta\rig$ of the average transverse momentum at different pseudorapidities.  Symbols are for results from realistic finite 
multiplicity events and lines are obtained from spectra integration (see caption of Fig. \ref{fig.rhocent}).} 
  \label{fig.rpt}
 \end{figure}
  
In Fig. \ref{fig.rpt} is shown the result of the calculation of the
 3-bin decorrelation measure $r_{p_T}(\eta)$ for three centralities (symbols).
 For centralities $30-40\%$
and $70-80\%$ the statistics is sufficient
 to indicate an increasing decorrelation
as the bin separation increases. Similar results are  obtained with 
oversampled events (lines), which gives a measure 
of the 3-bin decorrelation from 
collective flow only. We notice that the $p_T$-flow decorrelation 
in pseudorapidity  increases when going from central to peripheral collisions.

\section{Summary}
\label{sec.summary}

We investigate the event-by-event profile of the average 
transverse momentum of particles emitted in relativistic heavy-ion collisions.
The model used consist of relativistic viscous hydrodynamics. The initial 
conditions for the evolution come from 
a  Monte Carlo Glauber model  with quark degrees of freedom.
The model dynamics generates flow correlations from the collective expansion
 and non-flow correlations from resonance decays.
We compare the hydrodynamic simulation results to preliminary
 ALICE Collaboration data on the Pearson correlation coefficient  $b\lef [p_T]_F,[p_T]_B\rig$. We find a fair agreement for the magnitude of the 
correlation, and
qualitatively a similar centrality dependence.

We argue that the value 
 Pearson correlation coefficient  $b\lef [p_T]_F,[p_T]_B\rig$ measured in Pb+Pb collisions is dominated by statistical fluctuations of
 the average transverse momentum $[p_T]$ in finite multiplicity events.
This problem can be reduced by using a definition of the 
variance of $[p_T]$ that excludes self-correlations in the definition of the 
correlation coefficient. 
The proposed transverse flow-transverse flow correlation coefficient 
gives an estimate of the  correlation coefficient
for the collective transverse flow in two pseudorapidity bins.
The magnitude of this correlation is close to $1$ in our calculation,
several times larger than the standard Pearson correlation coefficient.
 
It would be interesting to measure in experiment the transverse
 flow-transverse flow
 correlation coefficient as a function of the pseudorapidity bin separation.
It would be a  measure of the correlations of the initial size of the  fireball 
in the longitudinal direction. The results could also point to additional 
$p_T$ correlations in rapidity arising during the dynamic evolution 
of the fireball. We give an estimate of the 3-bin measure of $p_T$ 
decorrelation in pseudorapidity defined as the ratio of  $p_T$ covariances 
calculated for different bin separations. With increasing bin separation
 $p_T$-$p_T$ correlation gets weaker. We find $5-10\%$ decorrelation over 
two units of $\eta$.

\begin{acknowledgments}

Research supported by the Polish Ministry of Science and Higher Education (MNiSW), by the National
Science Centre grant 2015/17/B/ST2/00101, as well as by PL-Grid Infrastructure. 

\end{acknowledgments}

\bibliography{../../hydr}

\end{document}